\documentclass{mn2e}
\usepackage{graphicx}
\usepackage{amssymb}
\footnotesize
\newdimen\minuswidth    
\setbox0=\hbox{$-$}
\minuswidth=\wd0
\catcode`@=\active
\def@{\kern\minuswidth}
\newdimen\digitwidth    
\setbox0=\hbox{\rm0}
\digitwidth=\wd0
\catcode`!=\active
\def!{\kern\digitwidth}
\normalsize
\title[Glitch activity in RRAT J1819$-$1458]
{Unusual glitch activity in the RRAT J1819$-$1458: an exhausted magnetar?}
\author[A. G. Lyne et al.]
{A. G. Lyne$^1$,
M. A. McLaughlin$^{2,3,4}$,
E. F. Keane$^1$,
M. Kramer$^{1,5}$,
C. M. Espinoza$^{1}$,
\newauthor
B. W. Stappers$^1$, N.~T. Palliyaguru$^{2}$ \& J. Miller$^{2}$
\\
$^1$ Jodrell Bank Centre for Astrophysics, 
School of Physics and Astronomy, The University of Manchester,
Manchester M13~9PL,UK\\
$^2$ Dept. of Physics, West Virginia University, Morgantown, WV 26506, USA \\
$^3$ National Radio Astronomy Observatory, Green Bank, WV 24944, USA\\
$^4$ Alfred P. Sloan Research Fellow \\
$^5$ MPI f{\"u}r Radioastronomie, Auf dem H{\"u}gel 69, 53121 Bonn, Germany \\}

\let\leqslant=\leq 
\date{}
\begin{document}
\maketitle
\newcommand{\setthebls}{
}
\setthebls

\begin{abstract}

We present an analysis of regular timing observations of the
high-magnetic-field Rotating Radio Transient (RRAT) J1819$-$1458
obtained using the 64-m Parkes and 76-m Lovell radio telescopes over
the past five years. During this time, the RRAT has suffered two
significant glitches with fractional frequency changes of
$0.6\times10^{-6}$ and $0.1\times10^{-6}$. Glitches of this magnitude
are a phenomenon displayed by both radio pulsars and
magnetars. However, the behaviour of J1819$-$1458 following these
glitches is quite different to that which follows glitches in other
neutron stars, since the glitch activity resulted in a significant
long-term net decrease in the slow-down rate.  If such glitches occur
every 30 years, the spin-down rate, and by inference the magnetic
dipole moment, will drop to zero on a timescale of a few thousand years.
There are also significant increases in the rate of pulse detection
and in the radio pulse energy immediately following the glitches.
\end{abstract}


\section{Introduction}

Rotating Radio Transients (RRATs) are sporadic radio sources from
which we detect bursts of emission of typically a few milliseconds
duration and which occur at intervals of between a few minutes and a
few hours.  Eleven of these elusive objects were discovered in a
single-pulse analysis of the Parkes Multibeam Pulsar Survey (McLaughlin et
al. 2006)\nocite{mll+06}, but they cannot be detected through their
periodicity using standard Fourier or periodogram techniques. However,
careful studies of the times of arrival of the bursts reveal
underlying periodicities which indicate that they are rotating neutron
stars. Since their discovery, a campaign of timing observations has
been implemented in order to monitor the rotational history of these
objects (Mclaughlin et al. 2009)\nocite{mlk+09}. Rotating once every
4.26 seconds, J1819$-$1458
is the most prolific bursting RRAT. At 1.4 GHz it exhibits $\sim3$
ms bursts with peak flux density as strong as $\sim10$ Jy with a burst
rate of $\sim20-30$ $\rm{hr^{-1}}$. The high burst rate enabled a
quick determination of a timing solution for this source and hence an
accurate position. The true nature of this source as a neutron star
was further illustrated by X-ray observations. These revealed a
thermal X-ray spectrum which is what is expected from a cooling
neutron star (Reynolds et al. 2006; Gaensler et al. 2007; McLaughlin
et al. 2007)\nocite{rbg+06, gmr+07, mrg+07}.

The discovery of RRATs raised interesting questions about the number
of neutron stars in the Galaxy. Initial estimates indicate that RRATs
may outnumber the normal radio pulsar population in the Galaxy by a
factor of 4 (McLaughlin et al. 2006)\nocite{mll+06}. It has been
argued that the observed Galactic supernova rate seems now to be
insufficient to account for the whole population of neutron stars
(Keane \& Kramer 2008)\nocite{keanekramer08}, which include radio
pulsars, millisecond pulsars, magnetars, RRATs, X-ray detected but
radio quiet Isolated Neutron Stars (INSs) and the Central Compact
Objects (CCOs), some of which are thought to be neutron
stars. Evolution between the various populations of neutron stars
could go some way towards resolving this problem.  Indeed, there is
evidence of an evolutionary trend connecting young pulsars with
magnetars (Lyne 2004)\nocite{lyn04}.  Among the observed properties
which suggest that the two populations may be linked are very low
braking indices \cite{lpgc96}, magnetar-like burst activity in one
young pulsar \cite{ks08,ggg+08}, low luminosity during quiescence in
various magnetars, resembling the X-ray emission of normal pulsars
\cite{aklm08}, and pulsed magnetar radio emission (Camilo et al. 2006;
Camilo et al. 2007)\nocite{crh+06,crhr07}.

Long-term timing of all these objects provides essential information needed
to understand any evolutionary relationships between the different
neutron star populations. Among the different trends and features
observed in long-term data we find glitches, which are discontinuous
events resulting in spin-up of the neutron star, and thought
to be caused by the transfer of angular momentum from the rotating
superfluid in the stellar interior to the outer crust.  Glitches
sometimes dominate the long-term spin evolution of pulsars (Lyne et
al. 1996)\nocite{lpgc96} and have been observed in numerous young
pulsars and several magnetars, appearing to be a normal phenomenon
among rotating neutron stars.

We describe the observations in section 2, before discussing the
unusual glitch activity observed in RRAT J1819$-$1458 in section 3. In
section 4 we discuss the implications of these results as they address
to the relationship between RRATs and other neutron stars.

\section{Observations}\label{sec:observations}

The observations reported here start with the discovery observation in
August 1998, followed by an intensive 5-year series of recordings
starting five years later.  The measurements were carried out using
the 64-m Parkes Telescope in Australia and the 76-m Lovell Telescope
at Jodrell Bank in the United Kingdom at a frequency of 1.4 GHz. At
Parkes, a dual-channel cryogenic receiver was used to receive
orthogonal linear polarisations. Each channel was processed in a
$512\times0.5$-MHz filterbank, added in polarisation pairs to provide
the total intensity, 1-bit sampled and recorded on tape. Observations
at Parkes were typically 30 min in duration. At Jodrell Bank, a
dual-channel cryogenic receiver was sensitive to the two hands of
circular polarisation which were each processed in a $64\times1$-MHz
filterbank before being added to produce total intensity, 1-bit
sampled and stored on tape. Observations at Jodrell Bank were 60 min
in duration. The longer observations at Jodrell Bank were due to the
lower sensitivity as a result of the narrower available bandwidth. We
note also that the presence of broadband impulsive radio frequency
interference hinders the identification of pulses from RRATs at both
observatories.  However, the quality of the observational data was
markedly improved by the use of the zero-DM filtering algorithm
(Eatough et al. 2009)\nocite{ekl09}.

Off-line, the data were dedispersed at the nominal dispersion measure
(DM) of the RRAT as well as at a number of adjacent DM values to aid
RFI discrimination.  The resulting time sequences were then searched
for significant single bursts of radiation, whose times of arrival
(TOAs) were recorded. We can compare the observed TOAs with a model of
the neutron star's rotational and astrometric parameters in a manner
similar to standard pulsar timing (see Lorimer \& Kramer
(2005)\nocite{lk05} or Lyne \& Graham-Smith (2006)\nocite{ls06}). The
only difference is that standard pulsar methods obtain TOAs from
integrated pulse profiles of as many as $10^2-10^4$ single pulses
whereas for RRATs individual pulses have a sufficiently high S/N 
that we can obtain TOAs from individual pulses (and indeed are
required to do so because of the sporadic emission). The differences
between the observed arrival times and the predicted arrival times are
known as the pulsar's ``timing residuals'' and the model parameters
(such as period, period derivative and sky position) are determined by
adjusting their values to minimise the rms value of the residuals.

\section{Results}\label{sec:results}

\subsection*{Tri-Modal Residuals}
\begin{figure}
\includegraphics[angle=0,width=8.5cm]{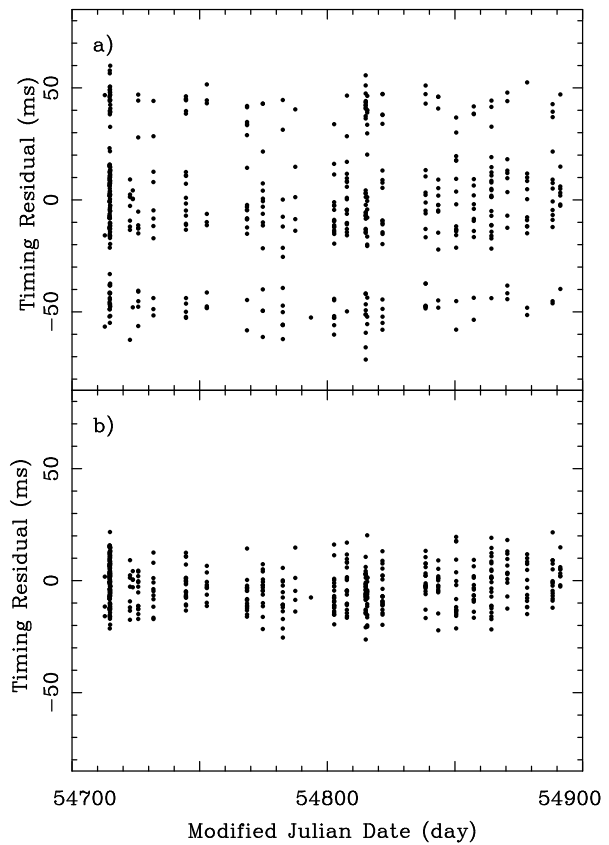}
\caption{a) The timing residuals from the TOAs of the individual
pulses from J1819$-$1458, relative to a simple slow-down model,
showing that they are located in 3 clearly identifiable bands which
are offset in pulse longitude by $\pm45$~ms.  b) the same residuals,
with the TOAs in the top band and the bottom band decreased and
increased by 45~ms respectively.  The rms of the residuals decreases
from 21.2~ms to 9.1~ms.  }\label{fig:resids}
\end{figure}

In Fig.~\ref{fig:resids}, we plot the timing residuals for all the
bursts received from J1819$-$1458 over the most recent 200 days
relative to a simple slow-down model which includes only the
spin frequency and its first derivative. The upper plot shows that
the bursts from J1819$-$1458 arrive over a longitude range spanning
approximately 120~ms and grouped within three longitude regions
separated by about 45~ms.

Fortunately, within a single 30-min or 60-min observation there is a 
sufficiently large number of bursts to allow each burst to be uniquely 
identified with one of the three bands. 
Those TOAs identified with the upper and lower bands are either
decreased by 45~ms or increased by 45~ms respectively and the
residuals recalculated. The lower diagram in Fig.~\ref{fig:resids}
shows the residuals resulting from this recalculation. As a result of
this procedure the rms value of the residuals is reduced by a factor
of $\sim2.5$ and the uncertainties in the fitted parameters are
similarly reduced. The timing analyses presented in this paper are
based upon all the TOAs modified in this fashion.

\begin{figure}
  \includegraphics[angle=-90,scale=0.35]{plots/residuals_22052009.ps}
  \hspace*{-4mm}
  \includegraphics[trim = 0mm 0mm 0mm 8mm, clip, angle=-90,scale=0.415]{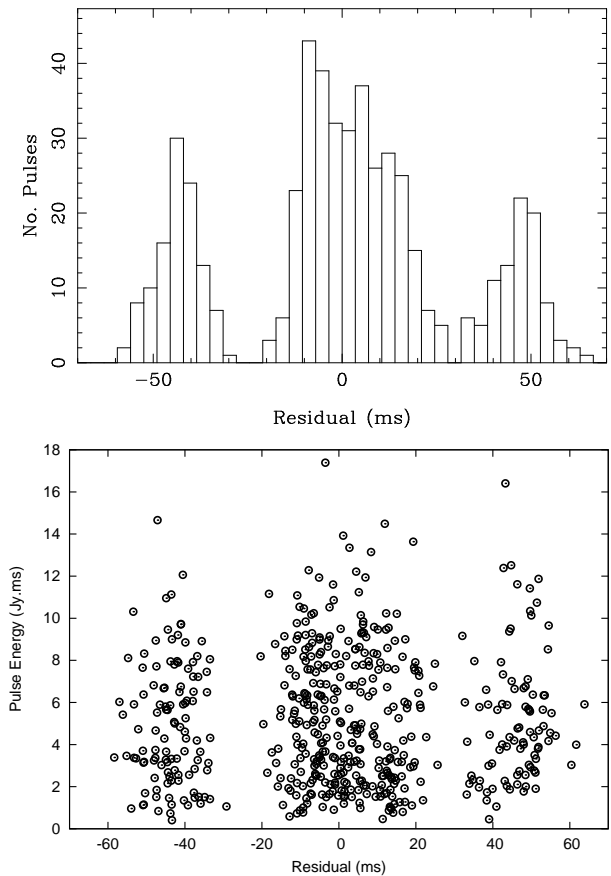}
    \caption{Top: Distribution of the timing residuals of bursts from
      RRAT J1819$-$1458 where the three bands seen in
      Fig.~\ref{fig:resids}a are clear. Bottom: A
      scatter plot of pulse energy versus
      timing residual.}\label{fig:band_intensities}
\end{figure}

\begin{figure}
  \includegraphics[angle=0,width=8.5cm]{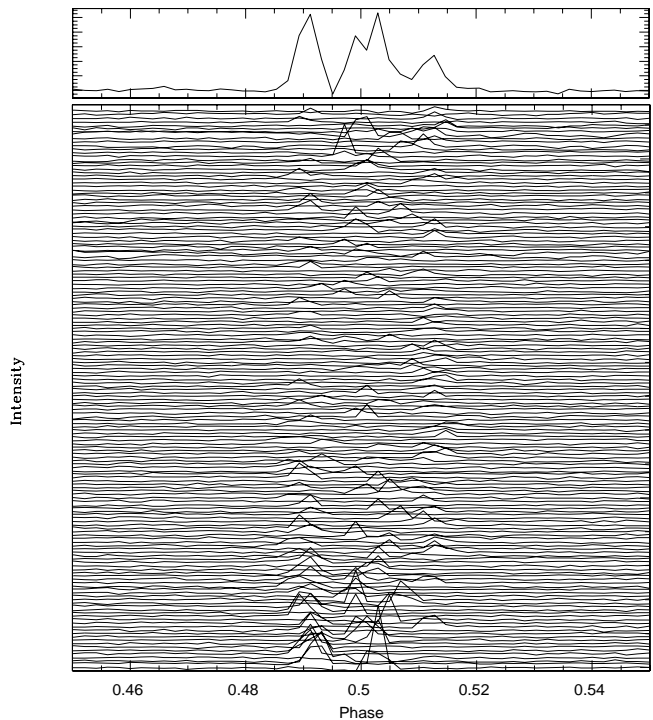} \caption{All
  165 single pulses detected during an eight-hr observation of
  J1819$-$1458 with Parkes at 1.4~GHz (bottom) along with the
  composite profile formed by summing all 165 pulses (top). The time
  intervals between the pulses vary.  }  \label{fig:composite}
\end{figure}

Most of the bursts ($\sim60\%$) come from the middle band, where also
the brightest bursts are found.  The top panel of
Fig.~\ref{fig:band_intensities} shows a collapsed histogram of the
timing residuals of the top panel of Fig.~\ref{fig:resids}. This is
essentially a probability distribution in rotational phase for the
bursts. The three-band structure is consistent with the
three-component profile reported by Karastergiou et
al. (2009)\nocite{khv+09}. In Fig.~\ref{fig:composite}, we show a
sequence of pulses detected during a single long Parkes observation,
along with the composite profile formed by summing all of these
individual pulses, which show anywhere from one to four components.
Such composite profiles vary from day to day due to the large degree
of pulse-to-pulse variation.

Esamdin et al. (2008)\nocite{ezy+08} reported the presence of only two
preferred longitude ranges, using the 25-m radio telescope at
Urumqi. However, because of the smaller size of that telescope, they
were only sensitive to the most intense bursts, those having flux
densities $\gtrsim3.4$ Jy, resulting in the observation of only 162
bursts in 94 hours of observation, or a burst rate of
1.7~hr$^{-1}$. Most of the bursts came from the middle band, with 18
from the early band and just 2 from the later band.  The higher
sensitivity of the larger telescopes used in the present experiment
allowed us to detect $\gtrsim500$ bursts from RRAT J1819$-$1458 in 27
hours of observation at rates of $\sim15-20$ $\rm{hr^{-1}}$ at Jodrell
Bank, and somewhat higher at Parkes. The bottom panel of
Fig.~\ref{fig:band_intensities} shows the intensity distributions of
bursts in the three bands. In determining pulse intensities, we used
the sky temperature model of Haslam et al. (1982)\nocite{hssw82}
extrapolated from 430-MHz to 1400-MHz assuming a spectral index of
$-2.6$ (Lawson et al. 1987)\nocite{lmop87}. Taking into account the
higher flux density limit of the Urumqi observations and the unknown
amplitude distribution within the bands at high flux density, the two
sets of results appear consistent.

The three-component profile is consistent with the tri-modal timing
residuals, and can be reconciled with a patchy emission beam with core
and conal components (e.g. Lyne \& Manchester 1988\nocite{lm88}). We
expect that many normal pulsars with profiles with multiple components
would exhibit similar bi- or tri-modal residuals if timed through
their single pulses.

\subsection*{Glitches}
\begin{figure}
  \includegraphics[angle=0,width=8.5cm]{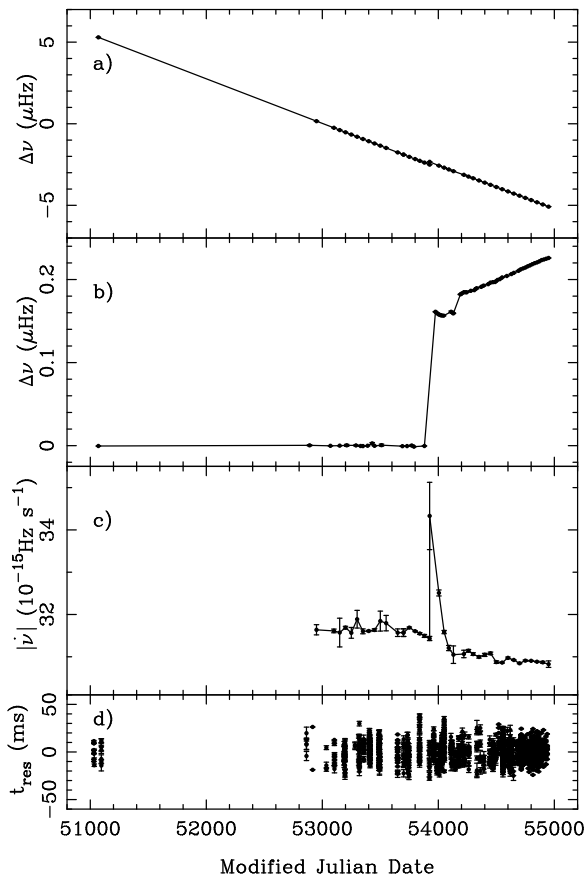}
  \caption{The frequency evolution of the RRAT J1819$-$1458 over a
  10-year period.  a) shows the secular slowdown in rotation rate of
  the neutron star, interrupted by a major glitch which is seen as a
  clear discontinuity at around MJD~53900. b) shows the frequency
  residuals relative to a simple slow-down model fitted to data
  between MJD 51000 and 53900 and reveals the presence of a second,
  smaller glitch, about 200 days later.  c) presents the variation in
  the magnitude of frequency derivative $|\dot\nu|$, showing a
  significant decrease in the rate of slowdown following the
  glitches. c) shows the timing residuals t$_{\rm res}$ relative to
  the rotational model given in Table~1.}\label{fig:freq}
\end{figure}

Fig.~\ref{fig:freq}a shows the evolution of the rotational frequency
over a $\sim$10-yr interval.  In the upper diagram, the steady
slowdown is interrupted at around MJD 53900 by a sudden spin-up. The
nature of this spin-up can be seen more clearly in
Fig.~\ref{fig:freq}b, which displays the frequency residuals relative
to a simple slow-down model fitted to the TOAs obtained prior to MJD
53925.  The spin-up is revealed as being due to two glitches, quite
close to one another, occurring at around MJD 53926 and MJD 54167,
followed by a period of steadily increasing frequency relative to the
model. The latter corresponds to a reduced magnitude of frequency
first derivative, which can be seen in Fig.~\ref{fig:freq}c following
a transient short-term increase.

\begin{table}
\begin{center}
\caption{The observed and derived rotational parameters of the RRAT
J1819$-$1458. }
\begin{tabular}{lc}
\hline
Timing parameters &  \\ \hline
Right Ascension $\alpha$  & $18^{\rm{h}}19^{\rm{m}}34^{\rm{s}}.173$ \\
Declination $\delta$  & $-14^\circ 58' 03''.57$  \\
Frequency $\nu$ (Hz) & 0.23456756350(2) \\
Frequency derivative $\dot\nu$ (s$^{-2}$) & $-31.647(1) \times 10^{-15}$ \\
Timing Epoch (MJD) & 54451.0 \\
Dispersion measure DM (cm$^{-3}$pc) & 196 \\
Timing data span (MJD) & 51031 -- 54938 \\
RMS timing residual $\sigma$ (msec) & 10.2 \\
\hline
Glitch 1 Parameters & \\ \hline
Epoch (MJD) & 53924.79(15) \\
Incremental $\Delta\nu$ (Hz) & 0.1380(6)$\times10^{-6}$ \\
Incremental $\Delta\dot\nu$ (s$^{-2}$) & 0.789(6)$\times10^{-15}$ \\
Decay $\delta\nu$ (Hz) & 0.0260(8)$\times10^{-6}$ \\
Decay timescale $\tau$ (days) & 167(6) \\
\hline
Glitch 2 Parameters & \\ \hline
Epoch (MJD) & 54168.6(8) \\
Incremental $\Delta\nu$ (Hz) & 0.0226(3)$\times10^{-6}$ \\ 
\hline
Derived Parameters & \\ \hline
Characteristic Age (kyr) & 120 \\
Surface Magnetic Field (G) & $50\times10^{12}$ \\ 
\hline
\end{tabular}
\end{center}
\label{tab:params}
\end{table}

In order to quantify these changes, the {\sc
TEMPO}\footnote{http://pulsar.princeton.edu/tempo} pulsar timing
package was used to model the slowdown of the pulsar and the two
glitches. It was found necessary to include steps in both $\nu$ and
$\dot\nu$ and a decaying exponential for the first glitch but only a
step in $\nu$ for the much smaller second glitch.  These parameters
are given in Table~1. For this timing analysis, we used the position
of J1819$-$1458 derived from Chandra X-ray observations of the source
(Rea et al. 2009)\nocite{}, which has errors of 0.2 arcsec in both
coordinates. The value of dispersion measure used was that given by
McLaughlin et al. (2006)\nocite{mll+06}.

Glitches usually display fractional increases in rotational frequency
$\Delta\nu/\nu$ which range between 10$^{-9}$ and 10$^{-5}$
(e.g. Lyne, Shemar \& Graham-Smith 2000)\nocite{lsg00}.  The two
glitches in J1819$-$1458 have $\Delta\nu/\nu=0.6\times10^{-6}$ and
$0.1\times10^{-6}$.  While these are not the largest glitches known
(c.f. Hobbs et al. 2002\nocite{hlj+02}) they are typical in fractional
size to those from many young pulsars (e.g. Zou et
al. 2008)\nocite{zwm+08} and magnetars \cite{mer08}.

\subsection*{Glitch-associated emission activity}

\begin{figure}
  \includegraphics[angle=0,width=8.5cm]{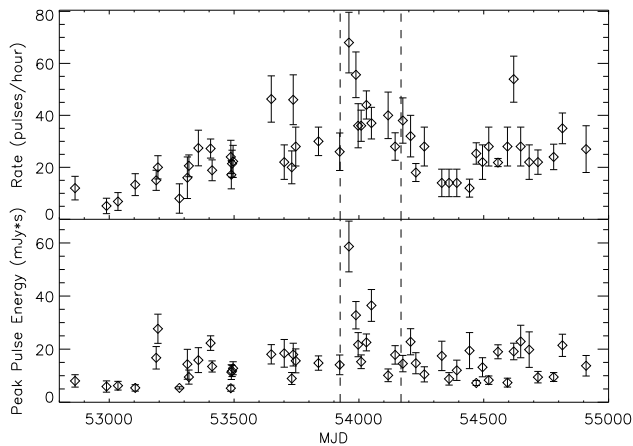}
  \caption{The variation of rate of pulse detection (top) and the peak
  pulse energy (bottom) with date for J1819$-$1458. The epochs of the
  two glitches are marked with dashed lines. As a control experiment,
  we find no dependence of the pulse detection rate on varying
  sensitivity due to the zenith angle of the observations.  }
  \label{fig:burstrate}
\end{figure}

In Fig.~\ref{fig:burstrate}, we show the rate of pulse detection (top)
and the peak pulse energy (bottom) for each observation as a function
of date for J1819$-$1458. For both quantities, there is a large
variation, with maximum values in the observation (on MJD~53960) which
immediately follows the first, and largest, glitch. On this date, the
mean pulse detection rate is 68$\pm$12 pulses/hour, 2.8 times the mean
and significant at the 3.5$\sigma$ level. The corresponding value of
the peak pulse energy is 58.7 mJy.s, which is 3.7 times the mean and
significant at the 4.7$\sigma$ level.  We note, however, that the high
burst rate on MJD 54619 is not associated with any obvious timing
abnormality. 


\section{Discussion}\label{sec:discussion}

\begin{figure}
  \includegraphics[angle=0,width=8.5cm]{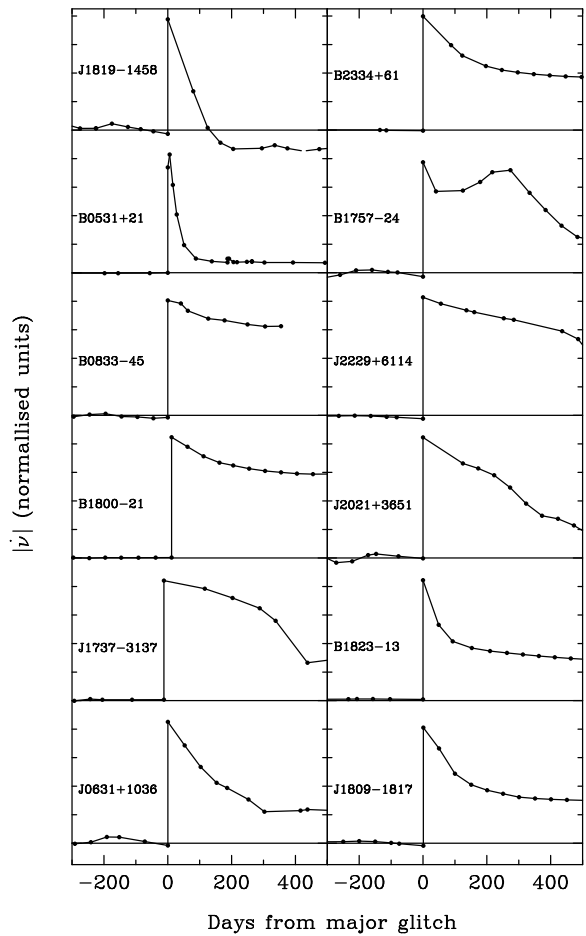} \caption{The
  changes in the magnitude of the first derivative of the rotational
  frequency $|\dot\nu|$ near the glitch from the RRAT J1819$-$1458 and
  near typical glitches in 11 pulsars. A second frequency derivative
  has been fitted to each pre-glitch data set and subtracted from the
  whole set. The scale of each diagram has been adjusted so that the
  step change in value at the glitch is the same.  Without exception,
  the final values of $|\dot\nu|$ for the pulsars are greater
  than before the glitches. In the case of
  J1819$-$1458, there is a net decrease in magnitude.}
  \label{fig:comparef1}
\end{figure}

While J1819$-$1458 exhibits many of the properties of normal,
relatively young radio pulsars, namely rapid spin-down and glitches,
the behaviour following the glitch activity is quite different. The
main effect is a long-term reduction in the magnitude of the slow-down
rate. This is unlike all observed glitches in radio pulsars and
magnetars, which always lead to a long-term increase in the rate of
slow-down.  Fig.~\ref{fig:comparef1} illustrates this by showing the
behaviour of slow-down rate for J1819$-$1458 and a selection of normal
young pulsars for a few hundred days around typical glitches.  The
behaviour of the Crab pulsar (B0531+21) shows the greatest similarity
to that of J1819$-$1458, each displaying a relatively short-term
transient before reaching a new asymptotic slow-down rate, although
the long-term net increments in slow-down rate clearly have different
signs.

\begin{figure}
  \includegraphics[angle=0,width=8.5cm]{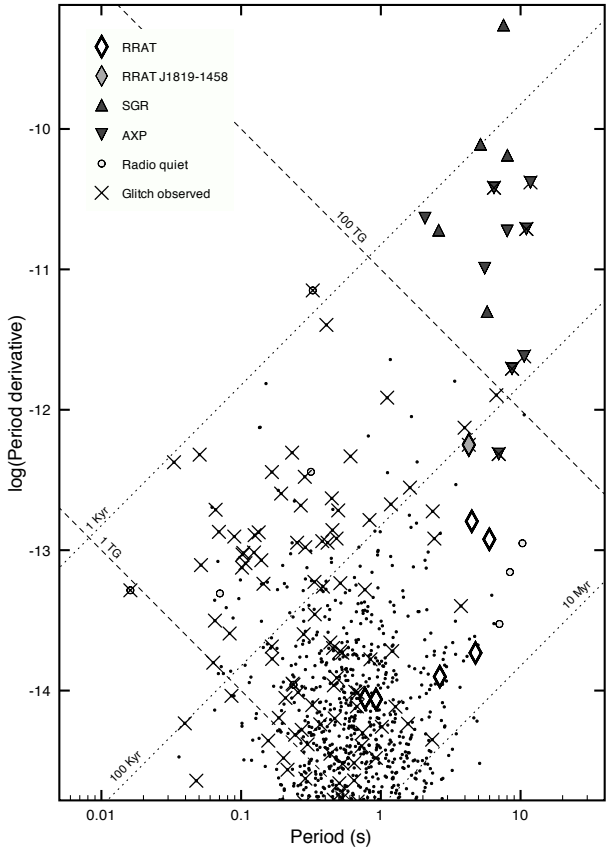} \caption{The
    $P$--$\dot{P}$ diagram for normal young pulsars and magnetars, as
    well as for the RRAT J1819$-$1458 (filled diamond) and other RRATs
    (open diamonds).} \label{fig:PPdot}
\end{figure}

Fig.~\ref{fig:PPdot} shows the $P$--$\dot{P}$ diagram for normal
pulsars, magnetars and other RRATs \cite{mlk+09}.  Objects with
constant dipolar magnetic field move towards the lower right-hand
corner of the diagram with a slope of $-1$.  Young normal pulsars are
observed to move with a slope of between $-1.0$ and +0.5. On this
diagram, glitches from normal pulsars such as those shown in
Fig.~\ref{fig:comparef1} result in a net increase in slow-down rate
and an upwards step in the $P$--$\dot{P}$ diagram.  On the other hand,
RRAT J1819$-$1458 stepped vertically downwards, towards smaller values
of $\dot{P}$. If this particular post-glitch behaviour is typical,
then the long-term effect of any glitches would be a secular movement
towards the bottom of the $P$--$\dot{P}$ diagram. If such glitches
were to occur every 30 years, then the slowdown rate would decay to
zero on a timescale of only a few thousand years.  Only larger time-span
observations will unveil the actual path of RRATs on the
$P$--$\dot{P}$ diagram. If the trend continues it could indicate that
the RRAT started off in the region of the diagram populated by the
magnetars.


Glitches are understood to be caused by the communication between a
superfluid component of the
neutron star interior and the crust. In the standard model, a neutron
star has a superfluid that is originally spinning faster than the
crust (see, e.g. Anderson \& Itoh 1975, Alpar \& Pines
1993\nocite{ai75,ap93}). This is because the magnetic dipole acts on
the crust and the coupled core, with the crustal superfluid rotating
independently. All neutron stars, regardless of their observational
manifestation, are expected to show this same structure. The angular
momentum resides on vortex lines, which are pinned to nuclei in the
crust. These lines suffer strong forces due to the mismatch between
the superfluid and crustal velocities. A glitch occurs when there is a
sudden unpinning of the vortex lines, and subsequent transfer of
angular momentum from the superfluid to the crust.  In a magnetar,
glitches could occur due to the high internal magnetic field that can
deform or crack the crust \cite{td96a}.

It could be that the strange glitch behavior of J1819$-$1458 is due to
a different mechanism for the unpinning, namely the deformation or
cracking of the crust due to the high magnetic field, as opposed to
the unpinning due to the angular velocity lag as suggested for normal
pulsar glitches.  While a magnetar glitch which resulted in a decrease in
the spin-down rate has not yet been observed, magnetar glitches
seem to show different characteristics to radio pulsar
glitches.  Gavriil et al. (2009)\nocite{gdk09} showed that AXP
4U~0142+61 suffered a glitch that resulted in a net spin-down. They
interpret this as being due to some regions of the superfluid
originally spinning slower than the crust, as opposed to faster in
normal rotation-powered pulsars. 

Magnetar glitches are occasionally associated with radiative events
and pulse profile and spectral changes (see, e.g. Dib et
al. 2008\nocite{dkg08b}), with no obvious relationship between the
size of the glitch and the extent of these changes.  The increase in
both burst rate and peak pulse energy immediately following the first
glitch in J1819$-$1458 is suggestive of an association. Both these
quantities seem to be sustained at a raised level for 100-200
days. Although there are some days with nearly as high a burst rate
that are not associated with any timing irregularity, the fact that
both burst rate and pulse energy are associated with the glitch
suggests that this is not just a statistical fluke.  More large
glitches must be detected for such correlations to be tested
robustly. In addition, X-ray observations immediately after the next
large glitch would be useful to search for magnetar-like bursts or
correlated pulse profile or spectral changes.

%
%

\section{Conclusion}

In this paper, we have demonstrated the difficulty of timing
pulsars with multiple components through their single pulses, and have
presented a simple algorithm for mitigating these effects.  

We have presented a timing analysis of the RRAT J1819-1458 over the
past 5\,yr and report the detection of two significant glitches.  The
magnitudes of these glitches are similar to those measured for radio
pulsars and magnetars, but the behaviour after the glitch is quite
different.  The glitches have resulted in a long-term net decrease in
the slow-down rate, corresponding to a downward vertical movement on
the $P-\dot{P}$ diagram. This jump in frequency derivative is
$\sim0.75(5)\times10^{-15}$ Hz $s^{-1}$, corresponding to a $\sim$1\%
decrease of the surface magnetic field of the neutron star.  If such
glitches are representative behaviour of this RRAT and were to occur
every 30 years, the spin-down rate, and by inference the magnetic
dipole moment, will drop to zero on a timescale of a few thousand
years. This implies that J1819$-$1458 may have begun its life in the
magnetar region of the diagram.

We have also briefly discussed the evolution of the burst rate and
peak flux density of pulses from J1819$-$1458 with time, and showed
that both peaked immediately after the large glitch. This is
tantalizing evidence for a similarity with magnetar glitches.

\section*{Acknowledgements}
We thank all the pulsar observers at Parkes for assistance with these
observations. MAM and NTP are supported by a WV EPSCOR grant.  EK
acknowledges the support of a Marie-Curie EST Fellowship with the FP6
Network ``ESTRELA'' under contract number MEST-CT-2005-19669. CME
thanks the support received from STFC and CONICYT through the
PPARC-Gemini fellowship PPA/S/G/2006/04449.


\end{document}